%% file: master.tex
\documentclass[journal,onecolumn]{IEEEtran}
\pdfoutput=1

\usepackage[utf8]{inputenc}
\usepackage[T1]{fontenc}
\usepackage{url}
\usepackage{ifthen}
\usepackage{cite}
\usepackage[cmex10]{amsmath} 

\usepackage{amsthm}

\newif\iffull
\fulltrue

\usepackage{todonotes}
\usepackage{amssymb}  
\newtheorem{theorem}{Theorem}
\newtheorem{example}{Example}
\newtheorem{definition}{Definition}

\theoremstyle{remark}
\newtheorem*{remark}{Remark}

\begin{document}
\title{Generalized Deduplication: \\ Bounds, Convergence, and Asymptotic Properties}
\author{%
  \IEEEauthorblockN{Rasmus Vestergaard, Qi Zhang, and Daniel E. Lucani\\}
  \IEEEauthorblockA{DIGIT and Department of Engineering\\
  				    Aarhus University, Denmark\\        
					\{rv, qz, daniel.lucani\}@eng.au.dk}
}

\maketitle
%
\input{sections/abstract}
\input{sections/introduction}
\input{sections/problemsetting}
\input{sections/bounds}

\input{sections/convergence}
\input{sections/numerical}
\input{sections/conclusion}

\newpage
\appendices
\section{A detailed example}\label{app:examples}
\input{appendix/example}

\section{Proof of theorem \ref{thm:generalizedBounds}}\label{app:genBounds}
\input{sections/proofGeneralizedBounds}

\section{Proof of theorem \ref{thm:deduplicationBounds}}\label{app:dedupBounds}
\input{sections/proofDeduplicationBounds}

\section{Proof of theorem \ref{thm:generalizedLimit}}\label{app:genLimits}
\input{sections/proofGeneralizedLimit}

\section{Proof of theorem \ref{thm:deduplicationLimit}}\label{app:dedupLimits}
\input{sections/proofDeduplicationLimit}

\input{ack}
\bibliographystyle{IEEEtran}
\bibliography{refs}
\end{document}

%% file: sections/abstract.tex


\begin{abstract}
We study a generalization of deduplication, which enables lossless deduplication of highly similar data and show that classic deduplication with fixed chunk length is a special case.
We provide bounds on the expected length of coded sequences for generalized deduplication and show that the coding has asymptotic near-entropy cost under the proposed source model.
More importantly, we show that generalized deduplication allows for multiple orders of magnitude faster convergence than classic deduplication.
This means that generalized deduplication can provide compression benefits much earlier than classic deduplication, which is key in practical systems.
Numerical examples demonstrate our results, showing that our lower bounds are achievable, and illustrating the potential gain of using the generalization over classic deduplication.
In fact, we show that even for a simple case of generalized deduplication, the gain in convergence speed is linear with the size of the data chunks.
\end{abstract}

%% file: sections/introduction.tex
\section{Introduction}
Deduplication~\cite{Xia2016} is a common practical compression technique in filesystems and other storage systems. It has been found to achieve significant space savings in several empirical studies for different workloads~\cite{El-Shimi2012,Meyer2012}.
Despite the practical importance, it has received little attention in the information theory community, with only Niesen's recent work analyzing its compression potential~\cite{Niesen2017}.
As more data is generated every year,
a thorough understanding of the fundamental limits of deduplication and similar techniques are of utmost importance. 

A significant shortcoming of deduplication is that near-identical files are not identified, and are considered as completely different files.
This can discourage the adoption of deduplication in some scenarios. An example is a network of Internet of Things (IoT) devices sensing an underlying process. Their measurements will be highly correlated, but may differ slightly due to spatial distance, measurement noise, and other factors.
Deduplication for data of this type can, to some extent, be enabled with a generalized view on deduplication%
. 
This \emph{generalized deduplication} allows near-identical chunks to be deduplicated, while still ensuring lossless reconstruction of the data.
The method has practical merits, and has been shown to achieve a compression of modelled sensor data in many cases where deduplication is unable to~\cite{Vestergaard2019a}.
Another instance is able to achieve a compression comparable to typical lossless compression methods for ECG data, while maintaining benefits from classic deduplication~\cite{Vestergaard2019b}.
This paper is a study of the theoretical properties of the technique, and it is shown how generalized deduplication compares to classic deduplication.
\subsection{Related work}
To our knowledge, Niesen presents the only previous information-theoretical analysis of deduplication~\cite{Niesen2017}. Niesen's work introduces a source model, formalizes deduplication approaches with chunks of both fixed-length and variable-length, and analyzes the performance of the approaches.
Our paper uses a similar strategy to analyze generalized deduplication.

The manner in which deduplication is presented will make it clear that it is similar to classic universal source coding techniques such as the LZ algorithms \cite{Ziv1977,Ziv1978}. In practice, the main difference between the methods is on the scale at which they operate. Deduplication attempts to identify large matching chunks (KB) on a global scale (GB to TB), whereas classic methods identify smaller amounts of redundancy (B) in a relatively small window (KB to MB).

The problem in deduplication is also similar to the problem of coding for sources with unknown alphabets~\cite{Orlitsky2004} or multiple alphabets~\cite{Aberg1997}. Such schemes attempt to identify the underlying source alphabet, and use this for universal compression, ideally approaching entropy regardless of the source's distribution. Deduplication can be seen as one such approach, compressing the source output by building a dictionary (alphabet) and replacing elements with a pointer to the dictionary.

\subsection{Contributions}
This paper provides a formal analysis of generalized deduplication and comparisons to classic deduplication, a special case. The main contributions are:

\paragraph{Bounds}
We present a simple model for generalized deduplication as a source coding technique. This model is used to derive upper and lower bounds on the expected length of encoded sequences. The potential gain of the generalization against the classic approach is bounded, quantifying the value of the generalization for data fitting the source structure.

\paragraph{Asymptotic behavior} 
We derive the asymptotic cost of generalized deduplication, showing that the method converges to as little as one bit more than the source entropy per chunk.
We analyze how fast this convergence happens, and show that the generalization allows for faster convergence.

\paragraph{Numerical results}
Concrete examples are used to show that the lower bounds are achievable. 
The generalization's potential for faster convergence and compression gain is easily visualized.

\iffull
Theorem proofs are deferred to the appendices.
\fi


%% file: sections/problemsetting.tex
\section{Problem Setting}
\input{sections/subsections/codingstrategies}
\input{sections/subsections/sourcemodel}

%% file: sections/subsections/codingstrategies.tex
\subsection{Generalized deduplication}
Generalized deduplication is now presented as a technique for source coding.  
In this paper, the technique operates on a randomly sampled binary sequence $s$, which consists of several \emph{chunks}. The chunks are restricted to have equal length, $n$ bits. 
The chunks in the sequence are a combination of a \emph{base} and a \emph{deviation}. The base is responsible for most of the chunk's information content, whereas the deviation is the (small) difference between the base and the chunk. This property of the data is important for the coding procedure.
Formally, the possible bases form a set $\mathcal{X}'$ and the deviations form a set $\mathcal{Y}$. These sets define the set of all potential chunks, $\mathcal{Z'} = \mathcal{X}'\oplus \mathcal{Y}$, i.e., the set arising from taking the symbol-wise exclusive-or of all bases in $\mathcal{X}'$ with all deviations in $\mathcal{Y}$.
The method requires identification of a minimum distance mapping $\phi: \mathcal{Z'} \rightarrow  \mathcal{X}'$, which will be used to identify a chunk's base.
The deviation can be found by comparing the chunk to its base.
The encoder and decoder must have prior knowledge of $\mathcal{X}'$ and $\mathcal{Y}$, which are used to determine the coded representations. 
These sets do not need to be stored explicitly.
The coders does not have prior knowledge of $\mathcal{Z}\subseteq \mathcal{Z}'$, which is the set chunks are generated from.
In particular, the coders does not know apriori which bases are the \emph{active} ones, which is some set $\mathcal{X}\subseteq \mathcal{X}'$, forming $\mathcal{Z} = \mathcal{X}\oplus \mathcal{Y}$. 

The presented algorithm encodes (decodes) a sequence in one pass, encoding (decoding) over a dictionary of previously encountered bases. 
In practical systems, data is structured in databases, since this enables independent and parallel access 
and higher speed. 
However, this paper follows the traditional source coding style of operating on a sequence, since this simplifies analysis.


\subsubsection*{Encoding}
The encoding procedure is initialized with an empty deduplication dictionary, $\mathcal{D}$.
To encode a sequence, it is processed sequentially, one chunk at a time. The mapping $\phi$ is applied to the chunk, identifying the base and the deviation.
The base is deduplicated against elements in $\mathcal{D}$. If it does not yet exist in the dictionary, it is added to the dictionary and this is indicated with a $1$ in the output sequence followed by the base itself.
If it already exists, this is indicated by a $0$ in the coded sequence followed by a pointer to the chunk's location in the dictionary, using $\lceil\log|\mathcal{D}|\rceil$ bits\footnote{All logarithms in this paper are to base $2$.}.
The deviation is added to the output sequence, following the base. It does not need to be represented in full, since knowing $\mathcal{Y}$ allows specification of a representation of $q\leq \lceil \log|\mathcal{Y}|\rceil$ bits.

\subsubsection*{Decoding}
The coded sequence is uniquely decodable. 
The decoding procedure is also initialized with an empty deduplication dictionary, $\mathcal{D}$.
Decoding happens one chunk at a time, parsing the sequence on the fly. 
If a $1$ is the first bit of a coded chunk, a base follows directly and is added to $\mathcal{D}$. 
On the other hand, if a $0$ occurs, the base was deduplicated, so it must already exist in $\mathcal{D}$, and is looked up based on the following pointer.
The coded deviation is expanded to its full representation.
Finally, the chunk can be reconstructed by combining the base and deviation. The reconstruction is added to the output sequence.
This is repeated until the coded sequence has been processed in its entirety.

\begin{remark}
The classic deduplication approach arises as an important special case.
It is obtained by considering each chunk as its own base, and thus there is no deviation. Formally, this means $\mathcal{Y}$ contains only the all-zero chunk of length $n$, so $\mathcal{X}' = \mathcal{Z}'$, and $\phi$ is the identity function.
\end{remark}

%% file: sections/subsections/sourcemodel.tex
\iffull
\else
\fi
\subsection{Source model}
A formal source model is now specified. 
All analysis in this paper uses this source structure.
Chunks will have a length of $n$ symbols, and are generated by a combination of two sources. Our analysis is restricted to binary symbols, so chunks are in the binary extension field~$\mathbb{Z}_2^n$.

The first source generates the active bases, and is denoted by $\mathcal{X} \subseteq \mathcal{X}'$. $\mathcal{X}'$ is a packing of $n$-dimensional spheres with radius $t$ in $\mathbb{Z}_2^n$.
The second source generates the deviations, and is denoted by $\mathcal{Y}$. This source consists of elements with low hamming weight, i.e., $\mathcal{Y} = \{v_i \in \mathbb{Z}_2^n : w(v_i) \leq t\}$ for the same $t$ as the packing.
This allows definition of the chunk source, $\mathcal{Z} = \mathcal{X} \oplus \mathcal{Y}$, which can be interpreted all points inside some spheres in $\mathbb{Z}_2^n$, where the spheres are centered at the bases from $\mathcal{X}$ and have radii $t$.
The fact that a sphere packing is used for $\mathcal{X}'$ implies that spheres are non-overlapping and, thus, $\mathbb{P}[Z=z] = \mathbb{P}[X=x]\cdot\mathbb{P}[Y = y]$ and $|\mathcal{Z}| = |\mathcal{X}||\mathcal{Y}|$.
We assume that chunks are drawn uniformly at random from $\mathcal{Z}$.

\begin{example}[Source construction]\label{ex:source}
Let $\mathcal{X'}$ be the set of codewords from the $(7,4)$ Hamming code and let $\mathcal{Y}$ consist of all binary vectors of Hamming weight at most $1$. Spheres of radii 1 cover the entire field, so $\mathcal{Z}' = \mathcal{X}' \oplus \mathcal{Y} = \mathbb{Z}_2^7$. In this example, let the base source have two active elements, e.g.,
\begin{align*}
\mathcal{X} = \{0000000, 1111111\},
\end{align*}
and $\mathcal{Z} = \mathcal{X}\oplus \mathcal{Y}$ then becomes
\iffull
\begin{align*}
\mathcal{Z} = \{&0000000, 0000001, 0000010, 0000100, 0001000, 0010000, 0100000, 1000000, \\
                &1111111, 1111110, 1111101, 1111011, 1110111, 1101111, 1011111, 0111111 \}
\end{align*}
\fi
with $|\mathcal{Z}| = |\mathcal{X}||\mathcal{Y}| = 16$. An optimal coding of this source uses $H(\mathcal{Z}) = \log|\mathcal{Z}| = 4$ bits per chunk.
The mapping $\phi:~\mathcal{Z}'\rightarrow~\mathcal{X'}$ (or $\mathcal{Z}\rightarrow\mathcal{X}$) can be derived from the decoding procedure for the Hamming code.
\end{example}

This source structure is a stylized model of the practical case where chunks tend to be similar, but not necessarily identical.
An example is a surveillance camera, continuously taking pictures of the same location.
The bases might then be the location in different lighting, and a change in some of the image's pixels can then be captured by the deviation.

%
\subsection{Coding a source}
%
Generalized deduplication has greater potential with large data sets and long chunks, yet a small example is useful to understand the method. An example is presented for the source of Example \ref{ex:source}. A step-by-step explanation of the encoding and decoding procedures is found in 
\iffull
appendix \ref{app:examples}.
\fi
We start with the simpler special case, classic deduplication.
\begin{example}[Deduplication]\label{ex:dedup}
Let $\mathcal{Z}$ be the source from Example \ref{ex:source}. Five chunks are chosen uniformly at random, and concatenated. This forms a sequence of $\ell(s) = 35$ bits\footnote{Delimiters are inserted between chunks for ease of reading; the coding and decoding procedures do not require this.}:
\begin{align*}
s = 0001000|0010000|0010000|1111110|0010000.
\end{align*}
Applying deduplication to this sequence results in:
\begin{equation*}
s_D = 1.0001000|1.0010000|0.1|1.1111110|0.01
\end{equation*}
where the final dictionary is $\{0001000, 0010000, 1111110\}$ and $\ell(s_D) = 29$ bits are used in total.
\end{example}

Let us now consider generalized deduplication. Full knowledge of $\mathcal{X}'$ and $\mathcal{Y}$ is available, and is used to determine the deviation representation and the minimum-distance mapping.

\begin{example}[Generalized deduplication]\label{ex:gen}
Consider again the sequence $s$ of Example \ref{ex:dedup}.
To apply generalized deduplication, a representation for the deviations is needed. As they are equiprobable $H(\mathcal{Y}) = \log |\mathcal{Y}| =  3$ bits, so $3$ bits is optimal for their representation. An optimal representation is
\begin{align*}
\{000 \leftarrow 0000000,~001 \leftarrow 0000001, \dots, 111 \leftarrow 1000000\}
\end{align*}
which in this special case is the syndrome representation of the (7,4) Hamming code. 
To compress the sequence, the minimum-distance mapping is applied to each chunk, identifying the closest base, which is a codeword of the Hamming code. The base is here represented in full, although it may easily be compressed to four bits since $\mathcal{X}'$ is known to be the set of~codewords from the (7,4) Hamming code. The result is:
\begin{equation*}
s_G = 1.0000000.100|0..101|0..101|1.1111111.001|0.0.101
\end{equation*}
where the final dictionary is $\{ 0000000, 1111111\}$ and $\ell(s_G) = 35$ bits are used.
\end{example}

Although in this limited example deduplication outperforms the generalization, our results show that this is not the case in general. In fact, the results show that there are significant benefits in convergence speed of using the generalized form. 

%% file: sections/bounds.tex
\section{Bounds}
In this section, the coded length of sequences is studied. 
Let $s$ be a random binary sequence of $C$ chunks of $n$ bits each, so $\ell(s) = C n$. The interesting metric is the expected coded length, given the length of the original sequence. 

\subsection{Bounds for coded sequence length for the generalization}
The expected length of the sequence after generalized deduplication is $R_G(C) = \mathbb{E}\left[\ell(s_G) | \ell(s) = Cn \right]$. This is decomposed as the sum of expected coded length of each chunk in $s$:
\iffull
\begin{IEEEeqnarray}{rCl}\label{eq:generalizedCost}
R_G(C) &= \sum\limits_{c=1}^{C} \mathbb{E}\left[ 1 + I\{ x_c \not\in \mathcal{D}^{c-1}\}(k + p) + I\{ x_c \in \mathcal{D}^{c-1}\}(l(\mathcal{D}^{c-1}) + p)  \right]
\end{IEEEeqnarray}
\fi
where $I\{\cdot\}$ is the indicator function%
, $\mathcal{D}^{c-1}$ is the dictionary after chunk $c-1$, $x_c$ is the base of chunk $c$, $l(\mathcal{D}^{c-1})$ is the number of bits needed to point to the dictionary, and finally $q$ is the number of bits used for representing the deviation. The base itself might be compressed to $k$ bits with $H(\mathcal{X}') \leq k \leq n$, since $\mathcal{X}'$ is known. 
Since chunks are drawn uniformly at random from $\mathcal{Z}$, this is equivalent to picking a base and a deviation uniformly at random from $\mathcal{X}$ and $\mathcal{Y}$. Thus,
\begin{IEEEeqnarray}{rCl}
\mathbb{P}[x_c \not\in \mathcal{D}^{c-1}] = (1 - |\mathcal{X}|^{-1})^{c-1} \triangleq p_{\mathcal{X}}(c).
\end{IEEEeqnarray}
%
We now state Theorem \ref{thm:generalizedBounds} bounding the expected length after generalized deduplication in the presented source model.
\begin{theorem}\label{thm:generalizedBounds}
The expected length of the generalized deduplication-encoded sequence from $C$ chunks of length $n$ is bounded as
\begin{IEEEeqnarray}{rCl}
\theta_L(C, \mathcal{X}, \mathcal{Y}) \leq R_G(C) \leq \theta_U(C, \mathcal{X}, \mathcal{Y}), \nonumber
\end{IEEEeqnarray}
where
\iffull
\begin{IEEEeqnarray}{rCl}
\theta_L(C, \mathcal{X}, \mathcal{Y}) =C(\log|\mathcal{Y}|+1) \label{eq:generalizedBoundsThm1} +\sum\limits_{c=1}^{C} \Big[ k p_{\mathcal{X}}(c)  + \left(1 - p_{\mathcal{X}}(c)\right) \log\left(|\mathcal{X}|\left(1 - p_{\mathcal{X}}(c)\right)\right) \Big]
\end{IEEEeqnarray}
and
\begin{IEEEeqnarray}{rCl}
\theta_U(C, \mathcal{X}, \mathcal{Y}) = C(\log\mathcal{|Y|}+3) \label{eq:generalizedBoundsThm2} + \sum\limits_{c=1}^{C} \Big[ k p_{\mathcal{X}}(c) + |\mathcal{X}|^{-1} \min\{ (c-1) \log(c-1) , |\mathcal{X}|\log|\mathcal{X}|\}\Big].  
\end{IEEEeqnarray}
\fi
\end{theorem}
\iffull
The proof of the theorem is reported in appendix \ref{app:genBounds}.
\else
\fi

\subsection{Bounds for coded sequence length for deduplication}
Classic deduplication is a special case which allows for a slightly closer upper bound, and is therefore treated separately.
The expected length of the sequence after deduplication is $R_D(C) = \mathbb{E}\left[\ell(s_D) | \ell(s) = Cn \right]$. With the previous notation, 
\iffull
\begin{IEEEeqnarray}{rCl}
R_D(C) &= \sum\limits_{c=1}^{C} \mathbb{E}\left[ 1 + I\{ z_c \not\in \mathcal{D}^{c-1}\}n + I\{ z_c \in \mathcal{D}^{c-1}\}l(\mathcal{D}^{c-1}) \right]\hspace*{-2.5pt},\quad
\end{IEEEeqnarray}
\fi
where $z_c$ is chunk $c$ itself, since it is now the base. This base cannot be compressed as before, so it needs $n$ bits.
\begin{theorem}\label{thm:deduplicationBounds}
The expected length of the deduplication-encoded sequence from $C$ chunks of length $n$ is bounded as
\begin{IEEEeqnarray}{rCl}
\theta_L(C, \mathcal{Z}, \boldsymbol{0}) \leq R_D(C) \leq \theta_U(C, \mathcal{Z},\boldsymbol{0}) - C\nonumber
\end{IEEEeqnarray}
where $\theta_L$ and $\theta_U$ are as in \eqref{eq:generalizedBoundsThm1} and \eqref{eq:generalizedBoundsThm2}
with $k=n$ since new chunks are represented with no compression, $\mathcal{Z}~=~\mathcal{X}~\oplus~\mathcal{Y}$ and $\boldsymbol{0}$ the set containing only the all-zero chunk of length $n$. 
\end{theorem} 
\iffull
The proof of the theorem is reported in appendix \ref{app:dedupBounds}.
\else
\fi
We illustrate the implications of Theorems \ref{thm:generalizedBounds} and \ref{thm:deduplicationBounds} through a numerical example in Section \ref{sec:simulations}.


\subsection{Bounds for the gain of generalized deduplication}
Theorems \ref{thm:generalizedBounds} and \ref{thm:deduplicationBounds} can be used to bound the expected gain from using generalized deduplication instead of deduplication.
\begin{definition}
The generalization ratio is 
\begin{IEEEeqnarray}{rCl}
G(C) = \frac{R_D(C)}{R_G(C)}. \nonumber
\end{IEEEeqnarray}
\end{definition}
The bounds of generalized deduplication from Theorem \ref{thm:generalizedBounds} and of deduplication from Theorem \ref{thm:deduplicationBounds} are used to loosely bound the generalization ratio as:
\begin{IEEEeqnarray}{rCl}
\frac{\theta_L(C, \mathcal{Z},\boldsymbol{0})}{\theta_U(C, \mathcal{X}, \mathcal{Y})} \leq G(C) \leq \frac{\theta_U(C, \mathcal{Z}, \boldsymbol{0}) - C}{\theta_L(C, \mathcal{X}, \mathcal{Y})}.
\end{IEEEeqnarray}
These bounds allow for a simple assessment of the expected gain in a specific scenario.

%% file: sections/convergence.tex
\section{Convergence}
\subsection{Asymptotic storage cost}
In this section, we provide theorems bounding the asymptotic coded length of a new chunk for generalized deduplication.
Let $\Delta R_G^{C}$ be the expected length of chunk $C$ when generalized deduplication is used, i.e., 
\begin{IEEEeqnarray}{rCl}
\Delta R_G^{C} = R_G(C) - R_G(C-1).
\end{IEEEeqnarray}
Then the asymptotic cost of generalized deduplication is bounded by Theorem \ref{thm:generalizedLimit}.
\begin{theorem}
Generalized deduplication has asymptotic cost 
\begin{IEEEeqnarray}{rCl}
H(\mathcal{Z}) + 1 \leq \Delta R_G^\infty \leq H(\mathcal{Z}) + 3 \nonumber
\end{IEEEeqnarray}
where $\mathcal{Z}$ is the set of potential chunks.\label{thm:generalizedLimit}
\end{theorem}
\iffull
The proof of the theorem is reported in appendix \ref{app:genLimits}.
\else
\fi
Generalized deduplication is thus asymptotically within one and three bits of the entropy of $\mathcal{Z}$. In practice, the method will operate on larger chunks with high entropy, so this overhead will be negligible.
Similarly, let $\Delta R_D^{C}$ be the expected length of chunk $C$ in classic deduplication:
\begin{IEEEeqnarray}{rCl}
\Delta R_D^{C} = R_D(C) - R_D(C-1).
\end{IEEEeqnarray}
For this special case, the closer upper bound in Theorem \ref{thm:deduplicationBounds} translates to a closer upper bound in asymptotic cost. 
\begin{theorem}
Classic deduplication has asymptotic cost 
\begin{IEEEeqnarray}{rCl}
H(\mathcal{Z}) + 1 \leq \Delta R_D^\infty \leq H(\mathcal{Z}) + 2 \nonumber
\end{IEEEeqnarray}
where $\mathcal{Z}$ is the set of potential chunks.\label{thm:deduplicationLimit}
\end{theorem}
\iffull
The proof of the theorem is reported in appendix \ref{app:dedupLimits}.
\else
\fi

\subsection{Rate of Convergence}
Now that it is established that generalized deduplication schemes converge to slightly more than the entropy of $\mathcal{Z}$, it is also important to quantify the speed of convergence.
Generalized deduplication should converge faster than deduplication in general, since the number of potential bases is smaller.
The generalization needs to identify $|\mathcal{X}|$ bases for convergence, whereas the classic approach requires $|\mathcal{X}||\mathcal{Y}| = |\mathcal{Z}|$ bases. Convergence of the classic approach thus requires identification of an additional factor of $|\mathcal{Y}|$ bases.
To formally analyze this, the following definition is needed~\cite[pp~12--13]{Suli2003}.
\begin{definition}
The rate of convergence of a sequence $\{a_1,a_2, ...\}$ converging to $\xi$ is
\begin{align*}
\mu = \lim_{i\rightarrow\infty} \left| \frac{a_{i+1} - \xi}{a_{i} - \xi}  \right|,
\end{align*}
with smaller values implying faster convergence.
\end{definition}

For generalized deduplication, convergence happens according to the convergence of $\lim_{c\rightarrow \infty}\mathbb{P}\left[ x_c \not\in \mathcal{D}^{c-1}\ \right] = 0$. This sequence has converged when $\mathcal{D}^{c-1} = \mathcal{X}$, and thus the summand in \eqref{eq:generalizedCost} is constant.
At this point $\Delta R_G$ remains constant, so it is sufficient to analyze the convergence of the sequence of probabilities. 
Thus,
\iffull
\begin{IEEEeqnarray}{rCl}
\mu_G = \lim_{c\rightarrow\infty} \left| \frac{\mathbb{P}\left[ x_{c+1} \not\in \mathcal{D}^{c}\ \right]}{\mathbb{P}\left[ x_c \not\in \mathcal{D}^{c-1}\ \right]}\right| 
= \lim_{c\rightarrow\infty} \frac{\left(1 - |\mathcal{X}|^{-1}\right)^{c}}{\left(1 - |\mathcal{X}|^{-1}\right)^{c-1}} 
= 1 - \frac{1}{|\mathcal{X}|}.
\end{IEEEeqnarray}
\fi
\begin{remark}
For the case of classic deduplication,
\begin{IEEEeqnarray}{rCl}
\mu_D = \mu_G\big| _{\mathcal{X} = \mathcal{Z}} &=& 1 - \frac{1}{|\mathcal{Z}|}.
\end{IEEEeqnarray}
Since $|\mathcal{Z}| \geq |\mathcal{X}| \Rightarrow \mu_D \geq \mu_G$. 
Thus, generalized deduplication will be able to converge faster. In fact, $|\mathcal{Z}| \gg |\mathcal{X}|$ even in simple cases. Both approaches exhibit \emph{linear} convergence~\cite{Suli2003}.
\end{remark}

%% file: sections/numerical.tex
\input{sections/objects/SimulatedDoubleFig}
\section{Numerical Results}\label{sec:simulations}
To visualize the results presented in the paper, a concrete example is considered.
The compression achieved by our method is compared to the compression achieved by zlib~\cite{zlib}, a well-known compression library implementing the popular DEFLATE algorithm~\cite[Section 6.25]{Salomon2010}, based on LZ77~\cite{Ziv1977} and Huffman coding~\cite{Huffman1952}.
\begin{example}
Let $\mathcal{X}'$ be the codewords of the $(31,26)$ Hamming code.
A subset $\mathcal{X}\subset X'$ with $|\mathcal{X}| = 8$ is chosen at random.
$\mathcal{Y}$ is the set of binary vectors of length $31$ with weight $1$ or less.
The resulting $\mathcal{Z}$ has $|\mathcal{Z}| = |\mathcal{X}||\mathcal{Y}| = 8\cdot 32 = 256$ elements.
To compare their performances, generalized deduplication, classic deduplication, and the DEFLATE algorithm are applied to $C$ chunks uniformly drawn from this source.

The upper and lower bounds of $R_{ \{D,G \} }(C)$ from Theorems \ref{thm:generalizedBounds} and \ref{thm:deduplicationBounds} are shown as dashed lines in Fig. \ref{fig:total}. 
The solid lines are simulated averages.
Our approach clearly outperforms the other approaches.
The performance of classic deduplication and the DEFLATE algorithm are, for this source, comparable while the deduplication dictionary is filling up.
At the end of the simulation, both classic deduplication and the generalization have a smaller representation than the one of the DEFLATE algorithm.
It is seen that both classic deduplication and the generalization are converging to the same slope.
The asymptotic slope comes from the asymptotic cost, $H(\mathcal{Z})+1$.
When both schemes have converged, a gap remains between the lines.
The gap remains constant, but eventually becomes negligible as $C\rightarrow \infty$.

The upper and lower bounds of $\Delta R_{ \{D,G \} }^C$ from Theorems \ref{thm:generalizedLimit} and \ref{thm:deduplicationLimit} are shown as dashed lines in Fig. \ref{fig:delta} as a function of the number of chunks, $C$.
The assessment of the convergence rate in the previous section is now visualized: The faster convergence of the generalization is easily seen.
Further, the solid line shows the average which is seen to approximate the lower bound.
This is because $|\mathcal{X}|, |\mathcal{Y}|$ and $|\mathcal{Z}|$ all are powers of two for this source, and thus no overhead (compared to the lower bounds) are used to represent neither bases, deviations, nor the entire chunks.
The DEFLATE algorithm is unable to approach the entropy, while the other approaches are.

\iffull
\begin{figure}[!t]
\centering
\hspace*{0.75cm}
\begin{minipage}{0.5\textwidth}
\hspace*{-1.5cm}
\centering
\includegraphics[scale=1]{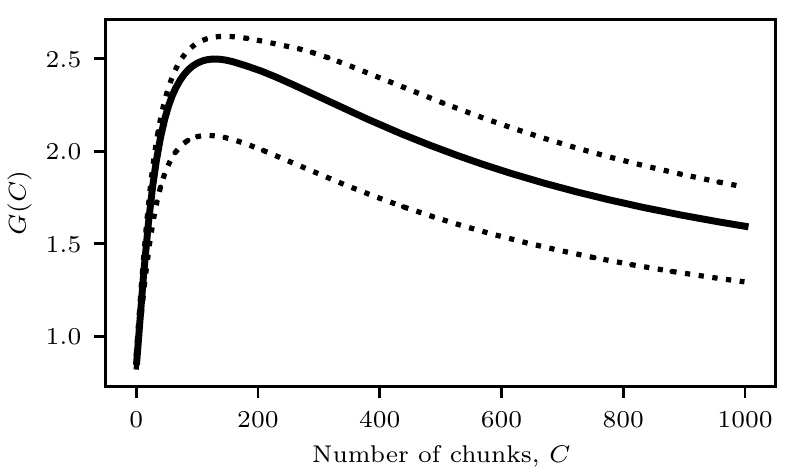}
\caption{Simulation and bounds for the generalization ratio, $G(C)$.\label{fig:gain}}
\end{minipage}
\end{figure}
\fi

The generalization ratio is shown in Fig. \ref{fig:gain}. For the first few chunks deduplication performs best, but this is quickly outweighed by the faster convergence of the generalization. 
The gain grows sharply until convergence of $\Delta R_G$, but slows down and then starts declining briefly thereafter. As the number of chunks goes to infinity, the ratio converges to $1$.
\end{example}

A general observation is that the maximum gain is achieved in the range where the generalization has converged, and classic deduplication is still far from converging.
It is also seen that, for the first few samples, the generalization performs slightly worse. This is caused by the convention to put the uncompressed base in the output. In reality, since $\mathcal{X}'$ is known, it is sufficient to use $\lceil\log|\mathcal{X}'|\rceil \leq n$ bits for each base. This will increase the gain slightly.

The vital advantage of generalized deduplication is the smaller number of bases, which causes more matches with fewer chunks.
\begin{example}
Let the mapping $\phi$ for generalized deduplication be defined through the (1023, 1013) Hamming code. Chunks must be $1023$ bits ($\approx 128$ B), and the potential bases $\mathcal{X'}$ are the codewords. $\mathcal{Y}$ is the set of binary vectors of length $1023$ with weight $1$ or less, so $|\mathcal{Y}| = 1024$. Thus $|\mathcal{Z}| = 1024|\mathcal{X}|$. The amount of bases in classic deduplication is three orders of magnitude greater than in the generalization.
\end{example}

By simulating sequences generated with longer chunks, it is clear that this increases the maximum generalization gain.
The convergence of deduplication is affected by an increase in $|\mathcal{Y}|$, which is unavoidable when changing the chunk size, unless the packing radius $t$ is also changed.
The generalization is oblivious of this, so its convergence will not be affected, and thus the potential gain increases. 
In practice, where limited amounts of data are available, this enables the generalization to achieve a significant gain in storage costs. 
Our simulations show that if $|\mathcal{X}|$ is fixed and the chunk length, $n$, is increased, then the maximum ratio, $\max_{C} G(C)$, increases \emph{linearly} as a function of the chunk length.
That is, the potential gain of using the generalization instead of classic deduplication increases linearly with the chunk length.
Fig. \ref{fig:differentconfigurations} shows the generalization ratio for three source configurations.
These simulations show a clear trend that when the number of unique chunks a source can output grows, then the potential advantage of using the generalization instead of classic deduplication becomes greater.

\iffull
\begin{figure}[!t]
\centering
\hspace*{0.75cm}
\begin{minipage}{0.5\textwidth}
\hspace*{-1.5cm}
\centering
\includegraphics[scale=1]{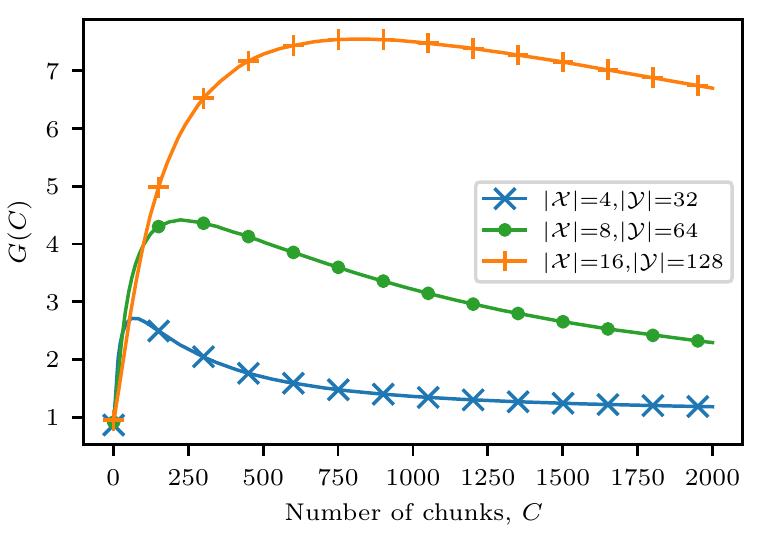}
\caption{Generalization ratio for different simulation configurations\label{fig:differentconfigurations}}
\end{minipage}
\end{figure}
\fi

%% file: sections/objects/SimulatedDoubleFig.tex
\iffull
\begin{figure}[!t]
\else 
\begin{figure*}[!ht]
\fi
\centering
\begin{minipage}{0.48\textwidth}
\centering
\includegraphics[scale=1]{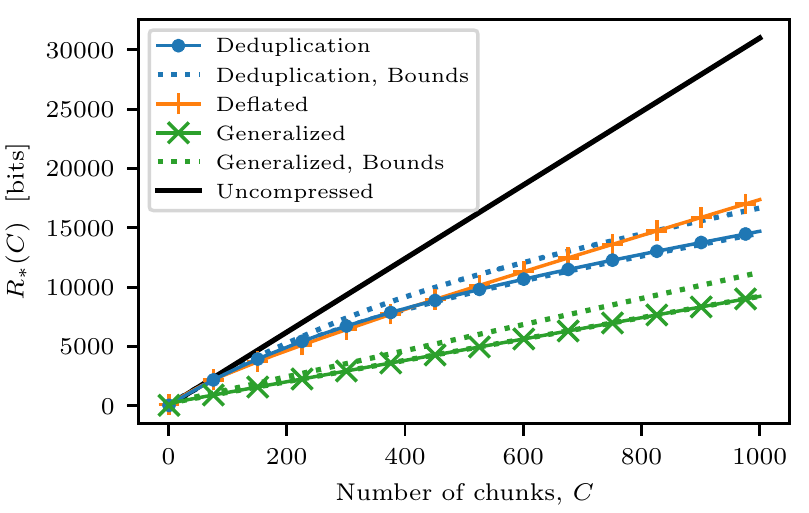}
\caption{Simulation and bounds for the expected sequence lengths, $R_D(C)$ and $R_G(C)$, and simulation for the DEFLATE algorithm.\label{fig:total}}
\end{minipage}
\hfill
\begin{minipage}{0.48\textwidth}
\centering
\includegraphics[scale=1]{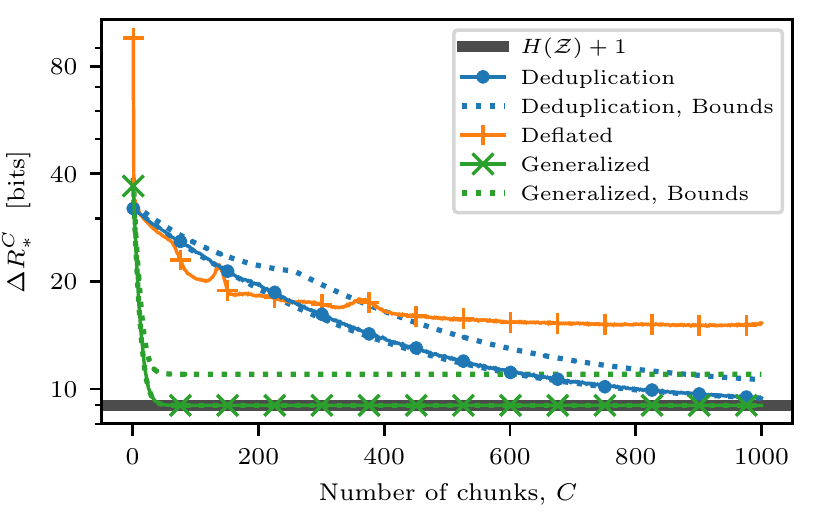}
\caption{Simulation and bounds for the expected number of bits per additional chunk, $\Delta R_D^C$, $\Delta R_G^C$ and the DEFLATE algorithm.\label{fig:delta}}
\end{minipage}
\iffull
\end{figure}
\else 
\end{figure*}
\fi

%% file: sections/conclusion.tex
\section{Conclusion}
The preceding sections present an information-theoretical analysis of generalized deduplication, which allows deduplication of near-identical data, and classic deduplication as a special case.
By analyzing a simple source model, we show that sources exist for which the advantages of the generalization are significant. 
Indeed, we show that generalized deduplication exhibits linear convergence with the number of data chunks.
In the limit each data chunk can be represented by at most 3 bits more than the entropy of the source, but our numerical results show that generalized deduplication can converge to the lower bound of 1 bit more than the entropy.
The advantage of generalizing deduplication manifests itself in the convergence. 
If the data has characteristics similar to our source model, then the generalization can converge to near-entropy costs with orders of magnitude less data than classic deduplication.
With an $m$-to-$1$ mapping $\phi$, a factor of $m$ fewer bases must be identified, creating a potential for improving compression in practice, where the amount of data will be limited.

The presented source model is somewhat stylized, and is not accurate for practical data sets.
An important next step is to lift the restriction of having data uniformly distributed over the spheres, which will enable a study of the method for general sources.
Indeed, our future work will address how to make the method more practical.
For instance, it is relatively simple to empirically model a chunk source, $\mathcal{Z}$, given concrete data, but this source must be carefully split into two underlying sources, the base source $\mathcal{X}$ and the deviation source $\mathcal{Y}$, in order to approximate the model and realize the potential of generalized deduplication.
We have studied some strategies for generalized deduplication from a more practical perspective~\cite{Vestergaard2019a,Vestergaard2019b}, but this task is not trivial in general.
We will continue with this work in the future.

%


%% file: appendix/example.tex
Assume that $\mathcal{X} = \{0000000, 1111111\}$, and let $\mathcal{Y} = \{v_i \in \mathbb{Z}_2^7 : w(v_i) \leq 1\}$. Let $\mathcal{Z} = \mathcal{X}\oplus \mathcal{Y}$. Draw 5 elements from $\mathcal{Z}$ i.i.d. uniformly. Assume that these elements are:
\begin{align*}
(0001000, 0010000, 0010000, 1111110, 0010000).
\end{align*}
The elements are then concatenated to a sequence:
\begin{equation*}
s = 0001000 0010000 0010000 1111110 0010000.
\end{equation*}

\subsection*{Classic Deduplication}
\subsubsection*{Encoding}
The encoding is initialized with an empty dictionary, $\mathcal{D}^0$. 
Since we know that chunks have length $7$, the sequence is split into chunks of that length:
\begin{equation*}k
s = 0001000|0010000|0010000|1111110|0010000.
\end{equation*}
Now, the chunks are handled sequentially. The first is $0001000$. This chunk is not in $\mathcal{D}^0$, so it is added to it. The new dictionary then is
\begin{equation*}
\mathcal{D}^1 = \{ 0001000\}
\end{equation*}
and the encoded sequence after the first chunk is formed by adding a $1$ (since we added the chunk to the dictionary) and then the chunk itself (the dot is only for easier visualization):
\begin{equation*}
s_D^1 = 1.0001000.
\end{equation*}
We then move to the next chunk, $0010000$, which is not in the $\mathcal{D}^1$. It is added, and a $1$ followed by the chunk is added to the encoded sequence:
\begin{align*}
\mathcal{D}^2 &= \{ 0001000, 0010000\}, \\
s_D^2 &= 1.0001000|1.0010000.
\end{align*}
The next element is $0010000$. This element is already in the dictionary, so it is not added again. For this reason, a $0$ is placed in the output sequence, followed by a pointer to the element in the dictionary using $\lceil\log|\mathcal{D}^2|\rceil = \lceil\log2\rceil = 1$ bit. Since the element is the second in the dictionary, it is represented by $1$:
\begin{align*}
\mathcal{D}^3 &= \mathcal{D}^2 = \{ 0001000, 0010000\}, \\
s_D^3 &= 1.0001000|1.0010000|0.1.
\end{align*}
The next element, $1111110$, is new. It is added to the dictionary, and the encoded sequence following a $1$:
\begin{align*}
\mathcal{D}^4 &= \{ 0001000, 0010000, 1111110\},  \\
s_D^4 &= 1.0001000|1.0010000|0.1|1.1111110.
\end{align*}
The final element is $0010000$, which already is in the dictionary. A pointer to the dictionary is therefore added to the encoding, following a $0$. The pointer now needs $\lceil\log|\mathcal{D}^4|\rceil = \lceil\log3\rceil = 2$ bits. Since the element is the second in the dictionary, it is represented as $01$. 
\begin{align*}
\mathcal{D}^5 &= \mathcal{D}^4 = \{ 0001000, 0010000, 1111110\},  \\
s_D^5 &= 1.0001000|1.0010000|0.1|1.1111110|0.01.
\end{align*}
All chunks are now encoded, and $s_D^5$ is output as $s_D$.

\subsubsection*{Decoding}
The encoding is initialized with an empty dictionary, $\mathcal{D}^0$. 
The sequence is processed sequentially. We start from 
\begin{align*}
s_D &= 10001000 10010000 01 11111110 001.
\end{align*}
The first bit is always a $1$, since the dictionary is empty. It is also known that chunks have length $7$. At first, the sequence can then be parsed as:
\begin{align*}
s_D &= 1.0001000|10010000 01 11111110 001.
\end{align*}
The first element can now be extracted and added to the dictionary. It is also added to the decoded sequence directly:
\begin{align*}
\mathcal{D}^1 &= \{ 0001000\}\\
s^1 &= 0001000.
\end{align*}
Since the inserted delimiter is followed by a $1$, it is known that the next chunk is also new. Therefore, a delimiter can be inserted $1+7 = 8$ bits after the first delimiter:
\begin{align*}
s_D &= 1.0001000|1.0010000|01 11111110 001.
\end{align*}
The chunk is added to the dictionary and the decoded sequence:
\begin{align*}
\mathcal{D}^2 &= \{ 0001000, 0010000\}, \\
s^2 &= 0001000|0010000.
\end{align*}
The new delimiter is followed by a $0$ flag this time. Therefore, the flag is followed by a pointer. Since $\lceil\log|\mathcal{D}^2|\rceil = \lceil\log 2\rceil = 1$, the flag is followed by a pointer of $1$ bit. A new delimiter can then be inserted:
\begin{align*}
s_D &= 1.0001000|1.0010000|0.1|11111110 001.
\end{align*}
The delimiter is followed by a $1$, which means that the second element in the dictionary should be added to the output sequence:
\begin{align*}
\mathcal{D}^3 &= \mathcal{D}^2 = \{ 0001000, 0010000\},  \\
s^3 &= 0001000|0010000|0010000.
\end{align*}
A $1$ follows the last delimiter, so a chunk follows directly. A new delimiter is inserted after the chunk:
\begin{align*}
s_D &= 1.0001000|1.0010000|0.1|1.1111110|001,
\end{align*}
and the chunk is inserted into the dictionary and the output, resulting in 
\begin{align*}
\mathcal{D}^4 &= \{ 0001000, 0010000, 1111110\},  \\
s^4 &= 0001000|0010000|0010000|1111110.
\end{align*}
Finally, a $0$ follows the delimiter. Since $\lceil\log|\mathcal{D}^4|\rceil = \lceil\log3\rceil = 2$, the two bits after the flag (which luckily is the rest of the sequence) points to an element in the dictionary. The value is $01$, so the second element in the dictionary should be added to the output sequence:
\begin{align*}
\mathcal{D}^5 &= \mathcal{D}^4 = \{ 0001000, 0010000, 1111110\},  \\
s^5 &= 0001000|0010000|0010000|1111110|0010000.
\end{align*}
The decoding is now complete, and $s^5$ is output as $\hat{s}$. Luckily $\hat{s} = s$, as expected. 
\subsection*{Generalized Deduplication}
\subsubsection*{Encoding}
As deviations are are drawn uniformly from $\mathcal{Y}$, $H(\mathcal{Y}) = \log |\mathcal{Y}| =  3$ bits. $3$ bits is thus optimal for their representation. An optimal representation is
\begin{align*}
\{&000 \leftarrow 0000000,~001 \leftarrow 0000001,~010\leftarrow 0000010,~011\leftarrow 0000100, \\
&100 \leftarrow 0001000,~101 \leftarrow 0010000,~110 \leftarrow 0100000,~111 \leftarrow 1000000\}.
\end{align*}
The encoding is initialized with an empty dictionary, $\mathcal{D}^0$. 
Since we know that chunks have length $7$, it is split into chunks of that length:
\begin{equation*}
s = 0001000|0010000|0010000|1111110|0010000.
\end{equation*}
The chunks are handled sequentially. The first is $0001000$. By applying the minimum distance mapping $\phi$ (decode and encode using that $\mathcal{X}'$ is the Hamming codewords), the base is found to be $0000000$. This base is not in $\mathcal{D}^0$, so it is added to it. In this example, we decide not to compress the base, but leave it in full size. The dictionary is then:
\begin{align*}
\mathcal{D}^1 &= \{ 0000000\}.
\end{align*}
Since the base was not in the dictionary, a $1$ is added to the sequence, and followed by the base. The deviation is the difference between the base, which in this case is $0001000$. The deviation is changed to the optimal representation. After the first chunk, the coded sequence is thus:
\begin{align*}
s_G^1 = 1.0000000.100.
\end{align*}
The next chunk is $0010000$. It also maps to the base $0000000$. A $0$ is added to the output sequence, followed by a pointer of $\lceil \log|\mathcal{D}^1|\rceil  = \lceil \log 1\rceil = 0$ bits pointing to the base. Since the base is the only element in the dictionary, no bits are needed to specify which one it is. The deviation is $0010000$, which is added in the optimal representation.
The dictionary and coded sequence thus becomes:
\begin{align*}
\mathcal{D}^2 &=\mathcal{D}^1, \\
s_G^2 &= 1.0000000.100|0..101.
\end{align*}
The next chunk is also $0010000$, and will get the same coded representation. Thus
\begin{align*}
\mathcal{D}^3 &=\mathcal{D}^2, \\
s_G^3 &= 1.0000000.100|0..101|0..101.
\end{align*}
This chunk, however, is followed by $1111110$. The nearest neighbor in $\mathcal{X}'$ (and $\mathcal{X}$) is $1111111$. This will thus be the base. The base is not in $\mathcal{D}^3$, so it is added to it, and
\begin{align*}
\mathcal{D}^4 &= \{ 0000000, 1111111\}.
\end{align*}
The deviation is found by comparing the chunk to the base, and is $0000001$. Changing this to the optimal representation, it is now possible to form the coded representation of the chunk. It is added to the encoding:
\begin{align*}
s_G^4 &= 1.0000000.100|0..101|0..101|1.1111111.001.
\end{align*}
Finally, the last chunk is $0010000$ again. The base is of course still $0000000$, and the deviation $0010000$. Although this base has been seen before, the representation in the output will be slightly different, since the dictionary has grown. Now $\lceil\log|\mathcal{D}^4|\rceil = 1$ bit is needed. The base is the first element in the dictionary, so it will be represented by a $0$:
\begin{align*}
\mathcal{D}^5 &= \mathcal{D}^4,\\
s_G^5 &= 1.0000000.100|0..101|0..101|1.1111111.001|0.0.101.
\end{align*}
The concludes the process, and $s_G^5$ is output as $s_G$. It is worth noting that already $\mathcal{D}^4 = \mathcal{X}$, and thus all subsequent chunks from $\mathcal{Z}$ will be represented with $5$ bits, one more than the entropy. This shows how the generalization can converge faster than classic deduplication. 

\subsubsection*{Decoding}
The encoding is initialized with an empty dictionary, $\mathcal{D}^0$. 
The sequence is processed sequentially. We start from 
\begin{align*}
s_G &= 1 0000000 100  0 101  0 101  1 1111111 001  0 0 101.
\end{align*}
The sequence starts with a $1$. This means that a base will follow the $1$ directly. The base is not compressed, so it has length $n=7$. The base is followed by a deviation represented with $3$ bits. This allows us to parse for the first chunk:
\begin{align*}
s_G &= 1.0000000.100|0 101  0 101  1 1111111 001  0 0 101.
\end{align*}
The base is added to the dictionary, so 
\begin{align*}
\mathcal{D}^1 &= \{ 0000000\},
\end{align*}
and the deviation is expanded to the full representation: $100\rightarrow 0001000$. The chunk is then reconstructed by combining the base and the deviation, using bitwise exclusive-or:
\begin{align*}
0000000 \oplus 0001000 = 0001000.
\end{align*}
This is the reconstructed chunk, which is added to the decoded sequence,
\begin{align*}
s^1 = 0001000.
\end{align*}
The next chunk has a $0$ flag, so the base is already in the dictionary. Since the dictionary has a single element only, $0$ bits are needed for the pointer. The deviation is as always $3$ bits. This allows the parsing of the second chunk to be made:
\begin{align*}
s_G &= 1.0000000.100|0..101|0 101  1 1111111 001  0 0 101.
\end{align*}
The base is then again $0000000$. The deviation is expanded: $101 \rightarrow 0010000$. These two are added, forming the new chunk:
\begin{align*}
0000000 \oplus 0010000 = 0010000,
\end{align*}
and this chunk is added to the output:
\begin{align*}
\mathcal{D}^2 &= \mathcal{D}^1, \\
s^2 &= 0001000|0010000.
\end{align*}
The third chunk starts with a $0$ too, so the base is indicated with $0$ bits, and is again the one already in the dictionary. The coded chunk is parsed as
\begin{align*}
s_G &= 1.0000000.100|0..101|0..101|1 1111111 001  0 0 101
\end{align*}
and is the same as the previous. The reconstruction is the same, so 
\begin{align*}
\mathcal{D}^3 &= \mathcal{D}^2, \\
s^3 &= 0001000|0010000|0010000.
\end{align*}
Now, the current last delimiter is followed by a $1$, so a new base of $7$ bits and a $3$-bit deviation follows. The parsing is
\begin{align*}
s_G &= 1.0000000.100|0..101|0..101|1.1111111.001|0 0 101.
\end{align*}
The base is $1111111$, and needs to be added to the dictionary:
\begin{align*}
\mathcal{D}^4 &= \{ 0000000, 1111111 \}.
\end{align*}
The deviation is then expanded, $001\rightarrow 0000001$. The base and deviation reconstructs the chunk:
\begin{align*}
1111111 \oplus 0000001 = 1111110,
\end{align*}
which is added to the output:
\begin{align*}
s^4 &= 0001000|0010000|0010000|1111110.
\end{align*}
The delimiter is now followed by a $0$, so the base is already in the dictionary. $\lceil\log|\mathcal{D}^4|\rceil = 1$ bit is used for the pointer, so the parsing is
\begin{align*}
s_G &= 1.0000000.100|0..101|0..101|1.1111111.001|0.0.101.
\end{align*}
The pointer is $0$, so the base is the first element in the dictionary, i.e., $0000000$. The deviation is $101 \rightarrow 0010000$, so the chunk can be combined to $0010000$. This means
\begin{align*}
\mathcal{D}^5 &= \mathcal{D}^4,\\
s^5 &= 0001000|0010000|0010000|1111110|0010000.
\end{align*}
The coded sequence is now fully decoded, and $\hat{s} = s^5$ is output. As expected, $\hat{s} = s$.

%% file: sections/proofGeneralizedBounds.tex
\begin{proof}
The structure of the source is such that drawing a chunk uniformly from $\mathcal{Z}$ is equivalent to drawing a base from $\mathcal{X}$ and a deviation from $\mathcal{Y}$. 
Since bases are drawn uniformly at random, the probability that the base of chunk $c$ is not already in the dictionary is 
\begin{equation}\label{eq:ProbNotSeenBefore}
\mathbb{P}\left[ x_c \not\in \mathcal{D}^{c-1}\ \right] = \left(1 - |\mathcal{X}|^{-1}\right)^{c-1}.
\end{equation}
The expected coded length can be bounded from below as:
\begin{align}
R_G(C) &= \sum\limits_{c=1}^{C} \mathbb{E}\left[ 1 + I\{ x_c \not\in \mathcal{D}^{c-1}\}(k + p) + I\{ x_c \in \mathcal{D}^{c-1}\}(l(\mathcal{D}^{c-1}) + p)  \right]  \nonumber\\
&\geq \sum\limits_{c=1}^{C} \mathbb{E}\left[ 1 + I\{ x_c \not\in \mathcal{D}^{c-1}\}(k + p) + I\{ x_c \in \mathcal{D}^{c-1}\}(\log|\mathcal{D}^{c-1}|+p)  \right] \label{eq:GLowerBound1}\\
  &= C(p+1) + \sum\limits_{c=1}^{C} \left( k\mathbb{P}\left[ x_c \not\in \mathcal{D}^{c-1}\ \right] + |\mathcal{X}|^{-1} \mathbb{E}\left[ |\mathcal{D}^{c-1}|\log|\mathcal{D}^{c-1}|\right]\right) \label{eq:GLowerBound2}\\
  &\geq C(\log|\mathcal{Y}|+1) + \sum\limits_{c=1}^{C} \left( k\mathbb{P}\left[ x_c \not\in \mathcal{D}^{c-1}\ \right] + |\mathcal{X}|^{-1} \mathbb{E}\left[ |\mathcal{D}^{c-1}|\log|\mathcal{D}^{c-1}|\right]\right)  \label{eq:GLowerBound3}\\
  &\geq C(\log|\mathcal{Y}|+1) + \sum\limits_{c=1}^{C} \left( k\mathbb{P}\left[ x_c \not\in \mathcal{D}^{c-1}\ \right] + |\mathcal{X}|^{-1} \mathbb{E}\left[ |\mathcal{D}^{c-1}|\right]\log\mathbb{E}\left[|\mathcal{D}^{c-1}|\right]\right)  \label{eq:GLowerBound4}\\
  &= C(\log|\mathcal{Y}|+1) + \sum\limits_{c=1}^{C} \left[ k \left(1 - |\mathcal{X}|^{-1}\right)^{c-1}  + \left(1 - \left(1 - |\mathcal{X}|^{-1}\right)^{c-1}\right) \log \left(|\mathcal{X}|\left(1 - \left( 1 - |\mathcal{X}|^{-1}\right)^{c-1}\right) \right)\right]  \label{eq:GLowerBound5}
\end{align}
where the inequality in \eqref{eq:GLowerBound1} follows from $\log|D^{c-1}| \leq l(\mathcal{D}^{c-1})$ because $l(\mathcal{D}^{c-1}) = \lceil \log|D^{c-1}|\rceil$.
The equality in \eqref{eq:GLowerBound2} uses that
\begin{align*}
\mathbb{E}\left[I\{ x_c \in \mathcal{D}^{c-1}\}\log|\mathcal{D}^{c-1}|\right] 
&= \mathbb{E}\left[ \mathbb{E} \left[ I\{ x_c \in \mathcal{D}^{c-1}\}\log|\mathcal{D}^{c-1}|~\big|~|\mathcal{D}^{c-1}| \right] \right] \\
&= \mathbb{E}\left[ \mathbb{P} \left[x_c \in \mathcal{D}^{c-1}~\big|~|\mathcal{D}^{c-1}|\right]\log|\mathcal{D}^{c-1}|  \right] \\
&= |\mathcal{X}|^{-1} \mathbb{E}\left[ |\mathcal{D}^{c-1}|\log|\mathcal{D}^{c-1}|\right].
\end{align*}
\eqref{eq:GLowerBound3} follows from $\log|\mathcal{Y}| \leq p$, since $p=\lceil \log|\mathcal{Y}|\rceil$.
The inequality in \eqref{eq:GLowerBound4} follows from Jensen's inequality, since $x\log x$ is a convex function.
Finally, the equality in \eqref{eq:GLowerBound5} comes from substituting (\ref{eq:ProbNotSeenBefore}) and the fact that
\begin{align*}
\mathbb{E}\left[|\mathcal{D}^{c-1}|\right] 
&= \sum\limits_{i=1}^{|\mathcal{X}|} \mathbb{P}\left[x_i \in\mathcal{D}^{c-1}|\right] \\
&= \sum\limits_{i=1}^{|\mathcal{X}|} 1 - \mathbb{P}\left[x_i \not\in\mathcal{D}^{c-1}|\right] \\
&= \sum\limits_{i=1}^{|\mathcal{X}|} 1 - \left(1 - |\mathcal{X}|^{-1}\right)^{c-1} \\
&= |\mathcal{X}|\left( 1- \left(1 - |\mathcal{X}|^{-1}\right)^{c-1}\right).
\end{align*}
Equivalently, the value can be bounded from above:
\begin{align}
R_G(C) &= \sum\limits_{c=1}^{C} \mathbb{E}\left[ 1 + I\{ x_c \not\in \mathcal{D}^{c-1}\}(k + p) + I\{ x_c \in \mathcal{D}^{c-1}\}(l(\mathcal{D}^{c-1}) + p)  \right]  \nonumber\\
&\leq \sum\limits_{c=1}^{C} \mathbb{E}\left[ 1 + I\{ x_c \not\in \mathcal{D}^{c-1}\}(k + p) + I\{ x_c \in \mathcal{D}^{c-1}\}(\log|\mathcal{D}^{c-1}|+1 + p)  \right] \label{eq:GUpperBound1}\\
  &\leq C(p+2) + \sum\limits_{c=1}^{C} \left( k\mathbb{P}\left[ x_c \not\in \mathcal{D}^{c-1}\ \right] + |\mathcal{X}|^{-1} \mathbb{E}\left[ |\mathcal{D}^{c-1}|\log|\mathcal{D}^{c-1}|\right]\right) \label{eq:GUpperBound2}\\
  &\leq C(\log\mathcal{|Y|}+3) + \sum\limits_{c=1}^{C} \left( k\mathbb{P}\left[ x_c \not\in \mathcal{D}^{c-1}\ \right] + |\mathcal{X}|^{-1} \mathbb{E}\left[ |\mathcal{D}^{c-1}|\log|\mathcal{D}^{c-1}|\right]\right) \label{eq:GUpperBound3}\\
& \leq C(\log\mathcal{|Y|}+3) + \sum\limits_{c=1}^{C} \left( k\mathbb{P}\left[ x_c \not\in \mathcal{D}^{c-1}\ \right] + |\mathcal{X}|^{-1} \min\{ (c-1) \log(c-1) , |X|\log|X|\}\right) \label{eq:GUpperBound4}\\
& = C(\log\mathcal{|Y|}+3) + \sum\limits_{c=1}^{C} \left( k \left(1 - |\mathcal{X}|^{-1}\right)^{c-1} + |\mathcal{X}|^{-1} \min\{ (c-1) \log(c-1) , |\mathcal{X}|\log|\mathcal{X}|\}\right) \label{eq:GUpperBound5}
\end{align}
where the inequality in \eqref{eq:GUpperBound1} follows from $l(\mathcal{D}^{c-1}) \leq \log|\mathcal{D}^{c-1}| + 1$ since $l(\mathcal{D}^{c-1}) = \lceil\log|\mathcal{D}^{c-1}|\rceil$, \eqref{eq:GUpperBound2} follows from the fact that $I\{\cdot\} \leq 1$. 
The inequality in \eqref{eq:GUpperBound3} is due to the encoding of the deviations, $p\leq \log|\mathcal{Y}| + 1$, since $p = \lceil\log|\mathcal{Y}|\rceil$.
The final inequality in \eqref{eq:GUpperBound4} follows from $|\mathcal{D}^{c-1}| \leq c-1$, and the fact that the maximum possible size of the dictionary is $|\mathcal{X}|$. Finally \eqref{eq:ProbNotSeenBefore} is substituted to get \eqref{eq:GUpperBound5}.
\end{proof}

%% file: sections/proofDeduplicationBounds.tex
\begin{proof}
The proof of the special case of deduplication naturally follows the same steps, but considers $\mathcal{Z} = \mathcal{X}$ and $\mathcal{Y}$ contains only the all-zero chunk. Because of this, deviations can be represented with exactly $0$ bits, so the step bounding their cost can be skipped. For completeness, the full proof is given. 
Since chunks are drawn from $\mathcal{Z}$ uniformly at random, the probability that chunk (=base) $c$ is not already in the dictionary is 
\begin{equation}\label{eq:ProbNotSeenBeforeZ}
\mathbb{P}\left[ z_c \not\in \mathcal{D}^{c-1}\ \right] = \left(1 - |\mathcal{Z}|^{-1}\right)^{c-1}.
\end{equation}
The expected coded length can be bounded from below as:
\begin{align}
R_D(C) &=\sum\limits_{c=1}^{C} \mathbb{E}\left[ 1 + I\{ z_c \not\in \mathcal{D}^{c-1}\}n + I\{ z_c \in \mathcal{D}^{c-1}\}l(\mathcal{D}^{c-1})  \right] \nonumber \\
&\geq \sum\limits_{c=1}^{C} \mathbb{E}\left[ 1 + I\{ z_c \not\in \mathcal{D}^{c-1}\}n + I\{ z_c \in \mathcal{D}^{c-1}\}\log|\mathcal{D}^{c-1}|  \right] \label{eq:DBelowBound1}\\
  &= C + \sum\limits_{c=1}^{C} \left( n\mathbb{P}\left[ z_c \not\in \mathcal{D}^{c-1}\ \right] + |\mathcal{Z}|^{-1} \mathbb{E}\left[ |\mathcal{D}^{c-1}|\log|\mathcal{D}^{c-1}|\right]\right) \label{eq:DBelowBound2}\\
  &\geq C + \sum\limits_{c=1}^{C} \left( n\mathbb{P}\left[ z_c \not\in \mathcal{D}^{c-1}\ \right] + |\mathcal{Z}|^{-1} \mathbb{E}\left[ |\mathcal{D}^{c-1}|\right]\log\mathbb{E}\left[|\mathcal{D}^{c-1}|\right]\right) \label{eq:DBelowBound3}\\
  &= C + \sum\limits_{c=1}^{C} \left[ n \left(1 - |\mathcal{Z}|^{-1}\right)^{c-1}  + \left(1 - \left(1 - |\mathcal{Z}|^{-1}\right)^{c-1}\right) \log \left(|\mathcal{Z}|\left(1 - \left(1 - |\mathcal{Z}|^{-1}\right)^{c-1}\right)\right) \right]\label{eq:DBelowBound4}
\end{align}
where the inequality in \eqref{eq:DBelowBound1} follows from $\log|D^{c-1}| \leq l(\mathcal{D}^{c-1})$ since $l(\mathcal{D}^{c-1}) = \lceil\log|D^{c-1}|\rceil$.
The equality in \eqref{eq:DBelowBound2} uses that
\begin{align*}
\mathbb{E}\left[I\{ z_c \in \mathcal{D}^{c-1}\}\log|\mathcal{D}^{c-1}|\right] 
&= \mathbb{E}\left[ \mathbb{E} \left[ I\{ z_c \in \mathcal{D}^{c-1}\}\log|\mathcal{D}^{c-1}|~\big|~|\mathcal{D}^{c-1}| \right] \right] \\
&= \mathbb{E}\left[ \mathbb{P} \left[z_c \in \mathcal{D}^{c-1}~\big|~|\mathcal{D}^{c-1}|\right]\log|\mathcal{D}^{c-1}|  \right] \\
&= |\mathcal{Z}|^{-1} \mathbb{E}\left[ |\mathcal{D}^{c-1}|\log|\mathcal{D}^{c-1}|\right].
\end{align*}
The inequality in \eqref{eq:DBelowBound3} follows from Jensen's inequality, since $x\log x$ is a convex function.
Finally, the equality in \eqref{eq:DBelowBound4} comes from substituting (\ref{eq:ProbNotSeenBeforeZ}) and the fact that
\begin{align*}
\mathbb{E}\left[|\mathcal{D}^{c-1}|\right] 
&= \sum\limits_{i=1}^{|\mathcal{Z}|} \mathbb{P}\left[z_i \in\mathcal{D}^{c-1}|\right] \\
&= \sum\limits_{i=1}^{|\mathcal{Z}|} 1 - \mathbb{P}\left[z_i \not\in\mathcal{D}^{c-1}|\right] \\
&= \sum\limits_{i=1}^{|\mathcal{Z}|} 1 - \left(1 - |\mathcal{Z}|^{-1}\right)^{c-1} \\
&= |\mathcal{Z}|\left( 1- \left(1 - |\mathcal{Z}|^{-1}\right)^{c-1}\right).
\end{align*}
The expected cost can also be bounded from above:
\begin{align}
R_D(C) &= \sum\limits_{c=1}^{C} \mathbb{E}\left[ 1 + I\{ z_c \not\in \mathcal{D}^{c-1}\}n + I\{ z_c \in \mathcal{D}^{c-1}\}l(\mathcal{D}^{c-1})  \right] \nonumber\\
&\leq \sum\limits_{c=1}^{C} \mathbb{E}\left[ 1 + I\{ z_c \not\in \mathcal{D}^{c-1}\}n + I\{ z_c \in \mathcal{D}^{c-1}\}(\log|\mathcal{D}^{c-1}|+1)  \right] \label{eq:DUpperBound1}\\
  &\leq 2C + \sum\limits_{c=1}^{C} \left( n\mathbb{P}\left[ z_c \not\in \mathcal{D}^{c-1}\ \right] + |\mathcal{Z}|^{-1} \mathbb{E}\left[ |\mathcal{D}^{c-1}|\log|\mathcal{D}^{c-1}|\right]\right)  \label{eq:DUpperBound2}\\
& \leq 2C + \sum\limits_{c=1}^{C} \left( n\mathbb{P}\left[ z_c \not\in \mathcal{D}^{c-1}\ \right] + |\mathcal{Z}|^{-1} \min\{ (c-1) \log(c-1) , |Z|\log|Z|\}\right)  \label{eq:DUpperBound3}\\
& = 2C + \sum\limits_{c=1}^{C} \left( n \left(1 - |\mathcal{Z}|^{-1}\right)^{c-1} + |\mathcal{Z}|^{-1} \min\{ (c-1) \log(c-1) , |\mathcal{Z}|\log|\mathcal{Z}|\}\right) \label{eq:DUpperBound4}
\end{align}
where the inequality in \eqref{eq:DUpperBound1} follows from $l(\mathcal{D}^{c-1}) \leq \log|\mathcal{D}^{c-1}| + 1$ since $l(\mathcal{D}^{c-1}) = \lceil\log|\mathcal{D}^{c-1}|\rceil$, \eqref{eq:DUpperBound2} follows from the fact that $I\{\cdot\} \leq 1$. The final inequality in \eqref{eq:DUpperBound3} follows from $|\mathcal{D}^{c-1}| \leq c-1$, and the fact that the maximum possible size of the dictionary is $|\mathcal{Z}|$. Finally, \eqref{eq:ProbNotSeenBeforeZ} is substituted to get \eqref{eq:DUpperBound4}.
\end{proof}

%% file: sections/proofGeneralizedLimit.tex
\begin{proof}
We use the bounds on the coded sequence length to to determine the asymptotic cost for each additional chunk.
First, the lower bound is proven, by assuming a best-case source that follows the lower bound on the coded sequence length.
From the derivation of the lower bound of $R_G(C)$, \eqref{eq:GLowerBound3} is restated:
\begin{align*}
R_G(C) \geq C(\log|\mathcal{Y}|+1) + \sum\limits_{c=1}^{C} \left( k\mathbb{P}\left[ x_c \not\in \mathcal{D}^{c-1}\ \right] + |\mathcal{X}|^{-1} \mathbb{E}\left[ |\mathcal{D}^{c-1}|\log|\mathcal{D}^{c-1}|\right]\right).
\end{align*}
By the definition of $\Delta R_G^{C+1} = R_G(C + 1) - R_G(C)$, a lower bound  on the expected coded length of chunk $C+1$ can be found from \eqref{eq:GLowerBound3}:
\begin{align*}
\Delta R^{C+1}_G \geq & (C+1)(\log|\mathcal{Y}|+1) + \sum\limits_{c=1}^{C+1} \left( k\mathbb{P}\left[ x_c \not\in \mathcal{D}^{c-1}\ \right] + |\mathcal{X}|^{-1} \mathbb{E}\left[ |\mathcal{D}^{c-1}|\log|\mathcal{D}^{c-1}|\right]\right)\\
& -\left( C(\log|\mathcal{Y}|+1) + \sum\limits_{c=1}^{C} \left( k\mathbb{P}\left[ x_c \not\in \mathcal{D}^{c-1}\ \right] + |\mathcal{X}|^{-1} \mathbb{E}\left[ |\mathcal{D}^{c-1}|\log|\mathcal{D}^{c-1}|\right]\right)\right)\\
= & \log|\mathcal{Y}| + 1 + k\mathbb{P}\left[ x_{C+1} \not\in \mathcal{D}^{C}\ \right] + |\mathcal{X}|^{-1} \mathbb{E}\left[ |\mathcal{D}^{C}|\log|\mathcal{D}^{C}|\right],
\end{align*}
and then the limit is 
\begin{align}
\Delta R_G^\infty &\geq \lim_{C \rightarrow \infty}\left( \log|\mathcal{Y}| + 1 + k\mathbb{P}\left[ x_{C+1} \not\in \mathcal{D}^{C}\ \right] + |\mathcal{X}|^{-1} \mathbb{E}\left[ |\mathcal{D}^{C}|\log|\mathcal{D}^{C}|\right]\right) \nonumber\\
&=\log|\mathcal{Y}| + 1 + \lim_{C \rightarrow \infty}k\mathbb{P}\left[ x_{C+1} \not\in \mathcal{D}^{C}\ \right] + \lim_{C \rightarrow \infty}|\mathcal{X}|^{-1} \mathbb{E}\left[ |\mathcal{D}^{C}|\log|\mathcal{D}^{C}|\right] \nonumber\\
&= \log|\mathcal{Y}| + 1  + |\mathcal{X}|^{-1} |\mathcal{X}|\log|\mathcal{X} \label{eq:GLimit3}|\\
&= \log|\mathcal{Y}| + 1  + \log|\mathcal{X}| \nonumber\\
&= H(\mathcal{Y}) + 1 + H(\mathcal{X}) \label{eq:GLimit5}\\
&= 1 + H(\mathcal{Z}), \label{eq:GLimit6}
\end{align}
where the equality in \eqref{eq:GLimit3} uses that all $x_c$ have non-zero probability, so the probability of not having any specific one in the dictionary goes to $0$, and the dictionary converges to the entire set of possible bases, $\mathcal{X}$. The fact that, by assumption, $\mathcal{Z}=\mathcal{X}\oplus\mathcal{Y}$ with non-overlapping spheres means that drawing chunks uniformly from $\mathcal{Z}$ is equivalent to drawing uniformly distributed elements from $\mathcal{X}$ and $\mathcal{Y}$, and so the relations $\log|\mathcal{Y}| = H(\mathcal{Y})$, $\log|\mathcal{X}| = H(\mathcal{X})$ and $H(\mathcal{Z}) = H(\mathcal{X}) + H(\mathcal{Y})$ holds. This is used for \eqref{eq:GLimit5} and \eqref{eq:GLimit6}.

Finally, a similar argument can be made for the upper bound, by assuming a worst-case source that follows the upper bound on the coded sequence length.
\eqref{eq:GUpperBound3} is restated from the earlier derivation of the upper bound on $R_G(C)$:
\begin{align*}
R_G(C) \leq C(\log\mathcal{|Y|}+3) + \sum\limits_{c=1}^{C} \left( k\mathbb{P}\left[ x_c \not\in \mathcal{D}^{c-1}\ \right] + |\mathcal{X}|^{-1} \mathbb{E}\left[ |\mathcal{D}^{c-1}|\log|\mathcal{D}^{c-1}|\right]\right)
\end{align*}
and, following the exact same steps as for the lower bound, the result is found to be 
\begin{align*}
\Delta R_G^\infty \leq 3 + H(\mathcal{Z}),
\end{align*}
concluding the proof.
\end{proof}

%% file: sections/proofDeduplicationLimit.tex
\begin{proof}
The proof of the special case of deduplication follows the same structure as the generalized version. 
First, the lower bound is proven, by assuming a best-case source that follows the lower bound on the coded sequence length.
 \eqref{eq:DBelowBound2} is restated from the derivation of the lower bound of $R_D(C)$.
\begin{align*}
R_D(C) &\geq C + \sum\limits_{c=1}^{C} \left( n\mathbb{P}\left[ z_c \not\in \mathcal{D}^{c-1}\ \right] + |\mathcal{Z}|^{-1} \mathbb{E}\left[ |\mathcal{D}^{c-1}|\log|\mathcal{D}^{c-1}|\right]\right).
\end{align*}
By the definition of $\Delta R_D^{C+1} = R_D(C + 1) - R_D(C )$, a lower bound  on the expected coded length of chunk $C+1$ can be found from \eqref{eq:DBelowBound2}:
\begin{align*}
\Delta R_D^{C+1} \geq &~C+1 + \sum\limits_{c=1}^{C+1} \left( n\mathbb{P}\left[ z_c \not\in \mathcal{D}^{c-1}\ \right] + |\mathcal{Z}|^{-1} \mathbb{E}\left[ |\mathcal{D}^{c-1}|\log|\mathcal{D}^{c-1}|\right]\right) \\
& -\left( C + \sum\limits_{c=1}^{C} \left( n\mathbb{P}\left[ z_c \not\in \mathcal{D}^{c-1}\ \right] + |\mathcal{Z}|^{-1} \mathbb{E}\left[ |\mathcal{D}^{c-1}|\log|\mathcal{D}^{c-1}|\right]\right)\right) \\
= &~  1 + n\mathbb{P}\left[ z_{C+1} \not\in \mathcal{D}^{C}\ \right] + |\mathcal{Z}|^{-1} \mathbb{E}\left[ |\mathcal{D}^{C}|\log|\mathcal{D}^{C}|\right].
\end{align*}
The limit can now be evaluated:
\begin{align}
\Delta R_D^\infty &\geq \lim_{C \rightarrow \infty} \left(1 + n\mathbb{P}\left[ z_{C+1} \not\in \mathcal{D}^{C}\ \right] + |\mathcal{Z}|^{-1} \mathbb{E}\left[ |\mathcal{D}^{C}|\log|\mathcal{D}^{C}|\right] \right) \nonumber\\
&= 1 + \lim_{C \rightarrow \infty} n\mathbb{P}\left[ z_{C+1} \not\in \mathcal{D}^{C}\ \right] + \lim_{C \rightarrow \infty}|\mathcal{Z}|^{-1} \mathbb{E}\left[ |\mathcal{D}^{C}|\log|\mathcal{D}^{C}|\right] \nonumber\\
&= 1 + |\mathcal{Z}|^{-1} |\mathcal{Z}|\log|\mathcal{Z}|\label{eq:DLimit3}\\
&= 1 + \log |\mathcal{Z}| \nonumber\\
&= 1 + H(\mathcal{Z}) \label{eq:DLimit5}
\end{align}
where \eqref{eq:DLimit3} uses that all $z_c$ has non-zero probability, so the probability of not having encountered any specific one before goes to $0$, and that the maximum size of the dictionary is $|\mathcal{Z}|$. Finally, $\log|\mathcal{Z}| = H(\mathcal{Z})$ in \eqref{eq:DLimit5} due to the uniform distribution.

An equivalent argument can be made for the upper bound, by assuming a worst-case source that follows the upper bound on the coded sequence length.
\eqref{eq:DUpperBound2}  is restated from the earlier derivation of the upper bound on $R_D(C)$:
\begin{align*}
R_G(C) &\leq 2C + \sum\limits_{c=1}^{C} \left( k\mathbb{P}\left[ z_c \not\in \mathcal{D}^{c-1}\ \right] + |\mathcal{Z}|^{-1} \mathbb{E}\left[ |\mathcal{D}^{c-1}|\log|\mathcal{D}^{c-1}|\right]\right).
\end{align*}
By repeating exactly the same steps as for the lower bound, the result is found to be
\begin{align*}
\Delta R_D^\infty \leq 2 + H(\mathcal{Z}),
\end{align*}
concluding the proof.
\end{proof}

%% file: ack.tex
\section*{Acknowledgments}
This work was partially financed by the SCALE-IoT project (Grant No. DFF-7026-00042B) granted by the Danish Council for Independent Research, the AUFF Starting Grant AUFF-2017-FLS-7-1, and Aarhus University's DIGIT Centre.